# Enhanced dual-beam excitation photoelectric detection of NV magnetic resonances in diamond

E. Bourgeois[a,b], E. Londero[c], K. Buczak[d], Y. Balasubramaniam[b], G. Wachter[d], J. Stursa[e], K. Dobes[f], F. Aumayr[f], M. Trupke[d], A. Gali[c,g], and M. Nesladek[a,b]

[a]IMOMEC division, IMEC, Wetenschapspark 1, B-3590 Diepenbeek, Belgium. [b]Institute for Materials Research (IMO), Hasselt University, Wetenschapspark 1, B-3590 Diepenbeek, Belgium. [c]Institute for Solid State Physics and Optics, Wigner Research Centre for Physics, Hungarian Academy of Sciences, PO Box 49, H-1525 Budapest, Hungary. [d]Vienna Center for Quantum Science and Technology, Atominstitut, TU Wien, 1020 Vienna, Austria. [e]Nuclear Physics Institute, v.v.i., ASCR, CZ-25068 Rez, Czech Republic. [f]Institute of Applied Physics, TU Wien, Wiedner Hauptstr. 8-10, 1040 Vienna, Austria. [g]Department of Atomic Physics, Budapest University of Technology and Economics, Budafoki út 8, H-1111 Budapest, Hungary.

Date: 2016-07-04

## Abstract

**The core issue for the implementation of the diamond NV centre qubits technology is the sensitive readout of NV spin state. We have recently demonstrated the photoelectric detection of NV magnetic resonances (PDMR), anticipated to be faster and more sensitive than optical detection (ODMR). Here we report on a PDMR contrast of 9 % - three times enhanced compared to previous work - on shallow N-implanted diamond. Based on ab-initio modelling, we demonstrate a novel one-photon ionization dual-beam PDMR protocol. We predict that this scheme is significantly less vulnerable to the influence of defects such as substitutional nitrogen.**

The negatively charged nitrogen-vacancy (NV⁻) centre in diamond has attracted particular attention as a room temperature solid state qubit (1) that can be read-out by optical detection of magnetic resonances (ODMR) (2). Numerous applications in the field of solid-state quantum information processing (3) and sensing (4) (5) (6) (7) (8) are being studied, including non-perturbing nanoscale magnetometry with single NV⁻ (9) and ultrasensitive magnetometry with NV⁻ ensembles (10).

We have recently demonstrated the photoelectric detection of NV⁻ electron spin magnetic resonances under green illumination (single-beam PDMR, or s-PDMR) (Fig. 1a), based on the electric detection of charge carriers promoted to the diamond conduction band (CB) by the ionization of NV⁻ and performed directly on a diamond chip equipped with electric contacts (11). Since the sensitivity of the magnetic resonance (MR) detection is inversely proportional to the product between the MR contrast and the root mean square of the number of detected photons $N_p$ or electrons $N_e$ (4) (12) (11), achieving a high photocurrent signal and a sufficient MR contrast is critical for further applications of PDMR in quantum technology. Using s-PDMR, we achieved a detection rate of ~ $10^7$ electrons.s⁻¹ per NV⁻ (N.A. of the objective: 0.95, green illumination: 3.4 mW, electric field: 2.4 $10^4$ V cm⁻¹), compared to 2 $10^4$ photons.s⁻¹ per NV⁻ using confocal ODMR detection. However, the one-photon ionization of single substitutional nitrogen ($N_s^0$) is one of the factors limiting the PDMR contrast, making a high green illumination power necessary to achieve a sufficient contribution of NV⁻ two-photon ionization to the total photocurrent (11). We observed that under blue illumination, the ionization of NV⁻ can be achieved by a more effective one-photon process, enhancing the proportion of the photocurrent associated with NV⁻ compared to $N_s^0$. Based on this idea, we developed the dual-beam excitation PDMR (d-PDMR) scheme (Fig. 1a), that is anticipated to lead to enhanced PDMR contrast in the case of samples with high $[N_s^0]/[NV^-]$ ratio and could therefore hold promise for the photoelectric readout of single NV⁻ spin (since the proportion of $N_s^0$ in the detection volume remains substantial even in the case of single NV⁻ centers contained in ultra-pure diamond) or for ultrasensitive diamond magnetometry with NV⁻ ensembles (for which irradiated type-Ib diamonds containing a high proportion of $N_s^0$ are used).

To determine the threshold for NV$^-$ and N$_s^0$ one-photon ionization, we first measured the photocurrent spectra of irradiated type-Ib diamonds. Based on the identification of the ionization bands, we designed the d-PDMR scheme, in which pulsed blue light (2.75 eV) directly promotes electrons from NV$^-$ ground state to the CB and converts the resultant NV$^0$ back to NV$^-$. Simultaneous CW green illumination (2.33 eV) independently controls the MR contrast by inducing spin-selective shelving transitions to NV$^-$ metastable state (13). By performing ab-initio calculations of N$_s^0$, NV$^-$ and NV$^0$ ionization cross-sections, we could explore the photo-physics related to the proposed scheme, relevant for achieving higher MR contrast and photocurrent signal.

The sample used for PDMR measurements (sample TP4) is an electronic grade type-IIa diamond implanted with $^{14}$N$^{4+}$ ions, resulting after annealing in the formation of a shallow NV$^-$ layer (density ~ 30 µm$^{-2}$, depth: 12 ± 4 nm). For photocurrent spectroscopy, an as-received type-Ib diamond plate (sample R, [N$_s^0$] ~ 160 ppm) was used as a reference, while two others were respectively proton- (sample A, [NV$^-$] ~ 35 ppm) and electron-irradiated (sample E2, [NV$^-$] ~ 20 ppm) and annealed. Coplanar electrodes with a distance of 100 µm (samples R, E2 and A) or 50 µm (sample TP4) were prepared on the surface of these diamond plates. The type-Ib samples were characterized by photoluminescence, FTIR and UV-visible absorption spectroscopy (see supplementary information).

To realize the d-PDMR scheme, a DC electric field (2.4 10$^4$ V cm$^{-1}$) is applied in between electrodes. A collimated blue (2.75 eV) laser beam, pulsed at 131 Hz, is focused in between electrodes onto the diamond surface using a 40X air objective (NA: 0.95, light spot diameter ~ 600 nm). CW green (2.33 eV) light produced by a linearly polarized Nd:YAG laser is combined with the blue beam using the same objective. The resulting photocurrent is pre-amplified and measured by a lock-in amplifier referenced to the blue light pulsing frequency, so that the photocurrent induced by CW green light does not contribute to the measured signal. The diamond chip is mounted on a circuit board equipped with microwave antennas (14). For photocurrent spectroscopy, monochromatic light (1 to 300 µW) pulsed at 12 Hz is focused onto the sample. At each photon energy, the photocurrent is pre-amplified and measured by lock-in amplification (photocurrent detected down to 3 fA). The photocurrent is normalized to the flux of incoming photons.

To gain insight into $NV^-$, $NV^0$ and $N_S^0$ photo-ionization mechanisms, we apply ab-initio Kohn-Sham density functional theory (DFT) calculations. In the photo-ionization process, an electron is excited from an in-gap defect level to the CB or from the valence band (VB) to an in-gap defect level. In our measurements, a bias voltage is applied to the sample, making the resultant electron or hole instantly leave the defects. The photo-ionization probability is then directly proportional to the absorption cross-section that depends on the imaginary part of the dielectric function related to the transition between the initial ground state and the final excited state. This process can be well approximated by the transition of a single electron from/to the in-gap defect level to/from the band edges, thus the imaginary part of the dielectric function can be calculated between the corresponding Kohn-Sham levels. In summary, the task is to calculate the excitation energies and the corresponding imaginary part of the dielectric function.

To this end, we calculate the lowest excitation energy that corresponds to the pure electronic transition [zero-phonon line (ZPL) energy] by the constraint DFT approach. Based on our previous studies (15) (16), we use a range-separated and screened hybrid density functional HSE06 (17) (18). We explicitly calculated the ZPL energies only for the band edges, and assumed that the excitations at higher energy follow the calculated band energies w.r.t. the band edge energy. The imaginary part of the dielectric function is calculated at the ground state geometry, following the Franck-Condon approximation. Optical transitions to the bands require accurate calculation of the electron density of states. Since HSE06 calculations with many k-points in the Brillouin-zone are computationally prohibitive, and given that PBE and HSE06 Kohn-Sham wave functions are very similar for $N_S$ and NV centres, we applied a generalised gradient approximated functional PBE to calculate their optical transition dipole moments (19). The defects were modelled in a 512-atom diamond supercell. Details about ab-initio calculations are presented in supplementary information.

The proposed d-PDMR scheme was tested on shallow NV⁻ ensembles implanted in electronic grade diamond (sample TP4). As a reference, s-PDMR was measured on the same sample. At a fixed microwave power of 1W, a maximum s-PDMR contrast of 8.9 ± 0.3 % was obtained (Fig. 1b), higher than the PDMR contrast previously observed on type-Ib and type-IIa diamond (11). Under green illumination the PDMR contrast is limited by the background photocurrent resulting from the ionization of $N_s^0$ (11). The enhanced s-PDMR contrast observed on sample TP4 can be explained by the higher contribution of the quadratic NV⁻ two-photon ionization to the total photocurrent (see supplementary Fig. 1), which is due to the confinement of the defects to the waist of the illumination beam – where the intensity is highest. By contrast, the illumination intensity in bulk samples decreases with depth, leading to a higher proportion of linear $N_s^0$ ionization in the total photocurrent.

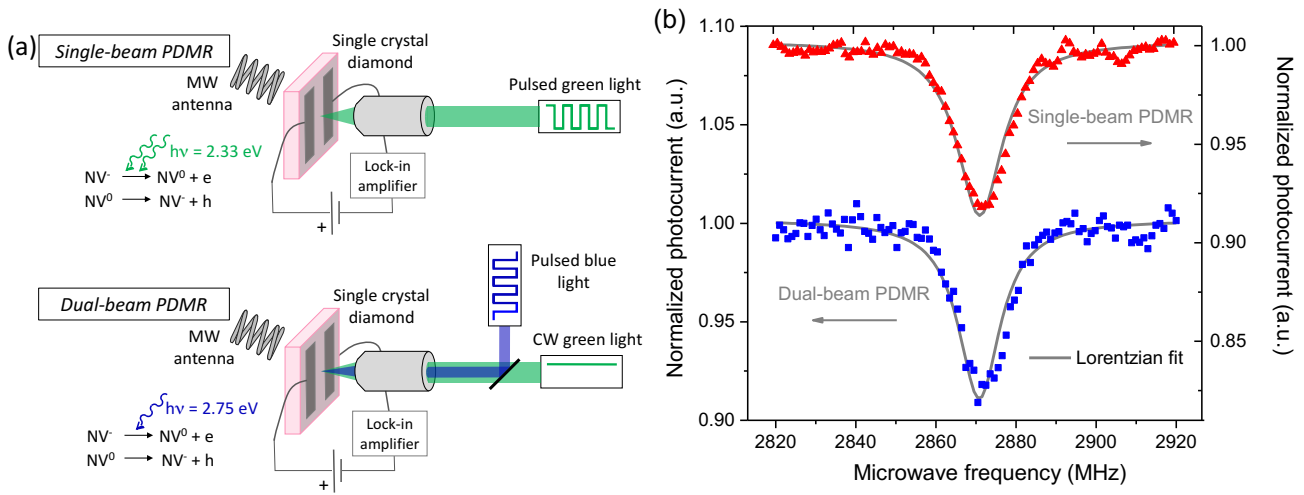

**Figure 1. (a)** Schematic diagram of the s-PDMR and d-PDMR schemes **(b)** Comparison between d-PDMR (pulsed blue excitation: 226 µW, CW green excitation: 9.1 mW) and s-PDMR (pulsed green excitation: 3 mW) spectra measured on shallow NV⁻ ensembles, in the conditions leading to maximum MR contrast (sample TP4, microwave power: 1W).

A maximum d-PDMR contrast of 9.0 ± 0.4 % is obtained on sample TP4 (Fig. 1b), close to the maximum s-PDMR contrast observed on the same sample at identical microwave power. Measurements of photocurrent as a function of green and blue light power on sample TP4 (supplementary Fig. 1b and 2b) show in addition that at identical power, blue one-photon excitation induces higher photocurrent than green excitation. For example, the photocurrent measured under 226 µW excitation (conditions leading to maximal d-PDMR contrast) is five times higher under blue (800 fA) than under green light (165 fA).

Although on shallow implanted NV$^-$ ensembles the MR contrasts obtained by d-PDMR and s-PDMR are similar, the d-PDMR scheme could potentially lead to higher contrast than s-PDMR in case of samples with high [$N_S^0$]/[NV$^-$] ratio, due to the lower contribution of $N_S^0$ ionization to the total photocurrent under blue illumination. Indeed, considering the green light power dependence of the photocurrent measured on type-Ib sample E2 (supplementary Fig. 1a), under 4 mW green excitation the two-photon ionization of NV$^-$ (quadratic fraction of the photocurrent) represents only ~ 1.5 % (0.6 pA) of the total detected photocurrent, while ab-initio calculations indicate that under 4 mW blue illumination ~ 30 % (0.6 nA) of the total photocurrent originates from the 1-photon ionization of NV$^-$ (see supplementary Fig. 8).

To explore the mechanism behind the d-PDMR scheme, we studied the energy dependence of NV and $N_S$ photo-ionization cross-sections by photocurrent spectroscopy. These measurements were performed on irradiated and annealed type-Ib diamonds, since a high density of defects allows to reach a high dynamic range of detected photocurrent, leading to a precise determination of photo-ionization thresholds.

The photocurrent spectrum measured on a type-Ib reference diamond (sample R, [$N_S^0$] ~ 160 ppm) can be observed in Fig. 2a. This spectrum displays a photo-ionization band with a threshold ionization energy of ~ 2.2 eV, obtained by fitting experimental data to Inkson's formula for the photo-ionization cross-section of deep defects (20). This band corresponds to the ionization of $N_S^0$ to $N_S^+$ (21) (22). Though its calculated (+|0) pure electronic charge transition level is at $E_C$-1.7 eV ($E_C$: CB minimum) the giant redistribution of position of the N and C atoms in the core of the defect upon ionization of $N_S^0$ results in a low ionization cross-section at this energy. Due to very strong electron-phonon interaction, a photo-ionization band emerges in the phonon sidebands around $E_C$-2.2 eV (see (23) and (24) for further discussion).

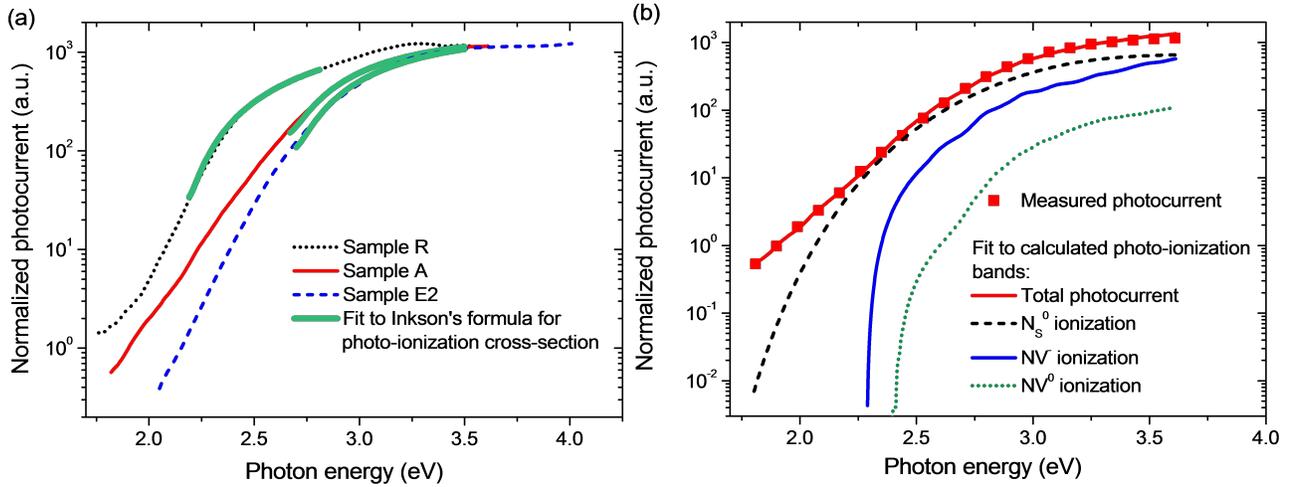

**Figure 2. Measured and calculated photo-ionization bands.** (a) Comparison between photocurrent spectra measured on reference type-Ib diamond [sample R, $E_i$ = 2.18(3) eV] and irradiated and annealed type-Ib diamonds [samples A, $E_i$ = 2.66(4) eV; sample E2, $E_i$ = 2.69(3) eV]. $E_i$: threshold ionization energy from Inkson's fitting. (b) Fitting of photocurrent measured on sample A using calculated ionization cross-sections. In the ab-initio calculation it was assumed that NV⁻, NV⁰ and $N_S^0$ ionization dominates the spectrum, with other parasitic defects contributing to a small extent to the spectrum in the low energy region (not shown).

Compared to non-irradiated diamond, the photocurrent spectra measured on proton- and electron-irradiated type-Ib diamonds (samples A and E2, in which ~ 10 % of $N_S$ defects are converted to NV⁻ centres) show a blue shift and the formation of an ionization band with threshold at ~ 2.7 eV (Fig. 2a). Photoluminescence, FTIR and optical absorption spectroscopy indicate that the dominant defects in these samples are $N_S$ and NV, with some additional spurious defects (possibly associated to Ni) in sample A (see supplementary note 3).

We calculated the ionization cross-sections of $N_S$ and NV defects as a function of the photo-excitation energy (see supplementary note 4) and compared the results with the photocurrent measurements (Fig. 2b for sample A, supplementary Fig. 8 for sample E2). In the simulation plots we set the experimental value of [$N_S^0$] and fit [NV⁰] and [NV⁻] to the experimental data. Using only these two fitting parameters we obtained [NV⁻] ≈ 31.4 ppm and [NV⁰] ≈ 1.0 ppm in sample A, in excellent agreement with the concentrations determined from photoluminescence measurements ([NV⁻] ≈ 34.0 ppm and [NV⁰] ≈ 1.1 ppm). Our ab-initio calculations predicted (25) that the (0|−) acceptor level of NV lies just in the middle of the diamond gap, at $E_C$-2.75 eV. Unlike $N_S$, NV⁰ and NV⁻ present very

similar geometries, thus pure electronic transitions dominate the ionization process. Photons with an energy above 2.7 eV can therefore ionize $NV^-$ to $NV^0$ by promoting an electron to the CB, but also convert $NV^0$ back to $NV^-$ by direct promotion of an electron from the VB. Calculations predict in addition that $NV^0$ has ~ 10 times larger ionization cross-section than $NV^-$ at 2.75 eV, implying a larger rate for the back-conversion than for the ionization of $NV^-$.

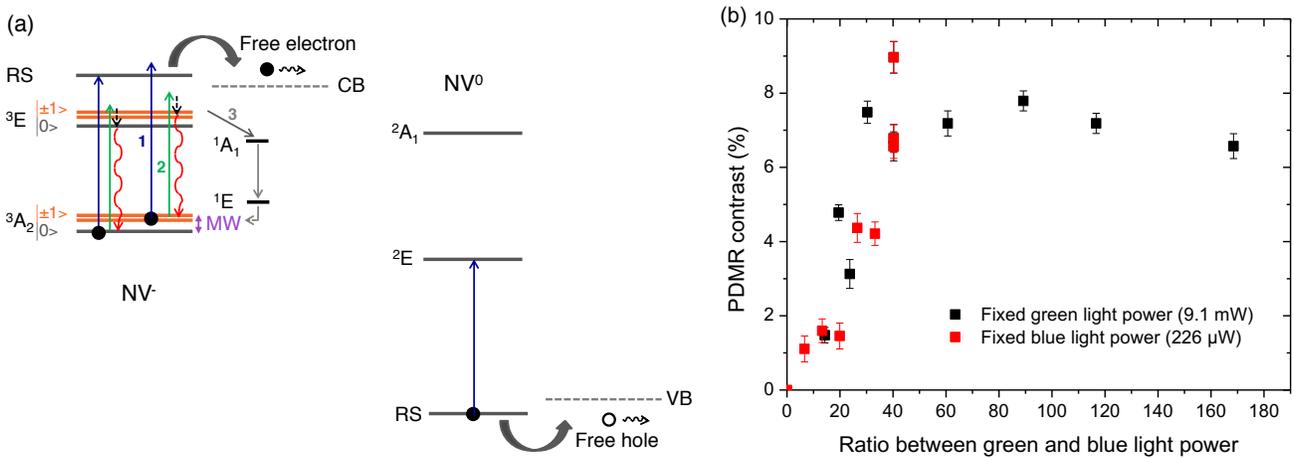

**Figure 3. d-PDMR on $NV^-$ centres. (a)** Schematic diagram of the d-PDMR mechanism (not to scale). Left: one-photon ionization of $NV^-$. Right: Back-conversion from $NV^0$ to $NV^-$. RS: resonant state. **(b)** d-PDMR contrast measured on shallow $NV^-$ ensembles as a function of the ratio $R_{GB}$ between the green and blue light powers (sample TP4). Error bars represent the standard errors of the fitting parameters.

Based on the results of photocurrent spectroscopy and ab-initio calculations, we explain the d-PDMR scheme as follows (Fig. 3a). Pulsed blue light (2.75 eV) promotes electrons from $NV^-$ triplet ground state $^3A_2$ to the CB by a one-photon process (transition 1), and induces also the one-photon back-conversion from $NV^0$ to $NV^-$ (transition 4). Simultaneous illumination by CW green laser light (2.33 eV) induces transitions from the ground state $^3A_2$ to the excited state $^3E$ (transition 2), followed by spin-selective non-radiative decay from the $|\pm1\rangle$ spin manifold of $^3E$ to the singlet state $^1A_1$ (transition 3) (13). From there, electrons fall to the metastable state $^1E$ (220 ns lifetime) (27). At resonant microwave frequency (2.87 GHz), these shelving transitions result in a temporary decrease in the occupation of $NV^-$ ground state, and thus to a decrease in the photocurrent associated with the one-photon ionization of NV⁻. Here we assume that the photo-ionization cross-section from the $^1E$ shelving state to the CB is low, although the metastable state $^1E$ state has been recently estimated to be located ~ 0.4 eV above NV⁻ ground state (i.e. ~ 2.3 eV below the CB) (28) and could therefore theoretically be ionized by 2.75 eV photons. However, the negative resonances observed in d-PDMR (Fig. 1b) indicate that the contribution of this process to the total photocurrent is significantly lower than the contribution of direct transitions from NV⁻ ground state to the CB.

Our d-PDMR model indicates that the relative rates between the direct ionization, back-conversion and shelving transitions to the metastable state, which can be controlled by varying the ratio $R_{GB}$ between the green and blue light powers, are dominantly responsible for the PDMR contrast. At a fix microwave power (1 W in the presented experiments), the PDMR contrasts obtained by varying the green light power at constant blue power and by varying the blue power at constant green power present a similar trend (Fig. 3b), which indicates that in the range of laser power considered here and for $R_{GB}$ < 40, the d-PDMR contrast rises with $R_{GB}$. The d-PDMR scheme allows thus an independent control of the photocurrent intensity (by the blue light power) and the MR contrast (by $R_{GB}$). The increase in the d-PDMR contrast with $R_{GB}$ (observed below $R_{GB}$ ≈ 40) can be explained by the transfer of an increased proportion of electrons initially in the |± 1> spin manifold to the metastable state. Above $R_{GB}$ ≈ 40, the contrast saturates and slightly decreases. It should be noted that this effect does not result from the saturation of the singlet state $^1E$, since it occurs when the blue light power is reduced at fixed green light power.

In conclusion, we demonstrated that a one-photon ionization scheme can be used for reading out the spin state of NV⁻. Based on this principle, we designed a novel photoelectric scheme for the detection of NV⁻ magnetic resonances, in which blue illumination induces the one-photon ionization of NV⁻ and converts NV⁰ back to NV⁻, while the MR contrast is independently controlled by CW green light. A maximal PDMR contrast of 9.0 % was obtained on shallow NV⁻ centres implanted in electronic grade diamond. The d-PDMR scheme is expected to be less sensitive than s-PDMR to background defects in diamond and to lead thus to enhanced MR contrast in the case of samples with high $[N_S^0]/[NV^-]$ ratio. This robust photoelectric detection scheme could therefore represent an important step toward the photoelectric readout of single NV$^-$ spin state and be used for the construction of diamond quantum opto-electronics devices with enhanced performances.

**Acknowledgements**

Support from EU (FP7 project DIADEMS, grant No. 611143) is acknowledged. A. G. acknowledges the Lendület program of the Hungarian Academy of Sciences. The authors thank A. Jarmola and D. Budker from the Department of Physics of the University of California (Berkeley, California) for the preparation of the electron-irradiated type-Ib diamond.

**Author contributions**

E.B. and K.B. performed the experiments. E.B. processed the PDMR data and performed the analysis. E.L. and A.G. carried out the ab-initio developments and calculations. Y.B. prepared electrodes on type-Ib diamond samples. G.W. designed and built the blue diode laser and designed the MW antennas. J.S. prepared the proton-irradiated type-Ib diamond. K.D. and F.A. performed the diamond ion implantation. M.T. proposed the use of implanted defects, designed the electrodes and assembled the device. M.N., M.T. and A.G. supervised the work. E.B., A.G. and M.N. wrote the manuscript. All authors discussed the results and commented on the manuscript.

**Supplementary Information** accompanies this paper.

**Competing financial interests:** The authors declare no competing financial interests.